\documentclass[11pt,a4paper]{article}
\usepackage{a4wide}
\usepackage[latin1]{inputenc}

\usepackage{amsmath,amscd,amssymb}
\usepackage[mathscr]{eucal}
\usepackage{float,graphicx,color}

\newcommand {\be}[1]{\begin{eqnarray} \mbox{$\label{#1}$}  }
\newcommand{\ee}{\end{eqnarray}}

\newcommand{\pref}[1]{(\ref{#1})}

\newcommand\ie {{\it i.e.}, }

\newcommand\etal {{\it et.al.  }}

\newcommand{\nn}{\nonumber\\}

\newcommand{\bs}{\boldsymbol}

\newcommand\half{\frac 1 2 }
\newcommand\halfsq{\frac 1 {\sqrt{2}} }

\newcommand{\dd}[2]{{d{#1}\over d{#2}}}

\renewcommand{\v}[1]{{\bf #1}}

\begin{document}

\title{Faster than light in a birefringent crystal}
\author{Tore Gunnar Halvorsen(a,b) and Jon Magne Leinaas(a) \footnote{toregha@gmail.com, j.m.leinaas@fys.uio.no}\\
(a) Department of Physics, University of Oslo,\\ P.O.
Box 1048 Blindern, 0316 Oslo, Norway\\
(b) Centre of Mathematics for Applications, University of Oslo,\\ P.O.
Box 1053 Blindern, 0316 Oslo, Norway}
\date{October 24, 2007}
\maketitle

\begin{abstract}
We examine the effect of superluminal signal propagation through a birefringent crystal, where the effect is not due to absorption or reflection, but to the filtration of a special polarization component.
We first examine the effect by a stationary phase analysis, with results consistent with those of an earlier analysis of the system. We supplement this analysis by considering the transit of a gaussian wave and find bounds for the validity of the stationary phase result. The propagation of the gaussian wave is illustrated by figures.
\end{abstract}

\section*{Introduction}
For wave propagation in a medium, group velocity is often regarded as being identical to signal velocity, \ie the velocity at which energy, and thereby information, is transferred. However, it has been known for a long time that group velocity, as usually defined, can exceed the speed of light without breaking causality \cite{sommerfeld14, brillouin14,brillouin}. In more recent years such superluminal propagation has been in the focus of interest, in particular since a series of different types of experiments now have been performed to demonstrate the effect. These experiments  have been done in media with anomalous dispersion and under conditions where superluminal tunneling times of single photons have been observed \cite{Chu82,Segard85,Steinberg93,Wang00}. 

Anomalous dispersion, and thereby superluminal group velocity, is usually obtained in systems with absorption or reflection. The relation between damping and group velocity is seen most directly in linear systems, where the Kronig-Kramers relations link the real part of the propagator to the imaginary part. The peculiar fact is that the KK relations, which on one hand reflects the condition of causality, on the other hand shows explicitly how superluminality arises in systems with anomalous dispersion.

Thus, even if group velocities in these systems may exceed the speed of light in vacuum, there is no conflict with causality. The apparent paradox, superluminality without breaking of causality, can be viewed as a consequence of the definition of group velocity, $v_
g=d\omega/dk$. In systems with anomalous dispersion $v_g$ typically varies rapidly with frequency $\omega$ in regions with large $v_g$, and to interpret this as the velocity of propagation therefore is meaningful only for signals that are sharply defined in frequency and therefore slowly varying in space and time. A signal with a sharper space profile will typically be rapidly deformed and damped so that superluminal propagation speed cannot be obtained over long distances. In fact for signals with a well defined front, the front velocity can never exceed the speed of light, and this fact is also the essence of the KK-relations.

Even if anomalous dispersion cannot occur in a transparent, {\em passive} medium, it is possible in an active medium with gain-assisted wave propagation. This has been shown both theoretically and experimentally \cite{Chiao93,Wang00}.  In such a case the signal is not damped, but the theoretical group velocity will also here only in an approximate sense represent the real speed of propagation for a signal with finite space extension. Thus, as discussed in \cite{Chiao93,Chiao96}  the signal propagation can be viewed as a pulse reshaping process. In this process the signal peak may move faster than light, but only for a limited time, since the signal front cannot be overtaken.

Some time ago D.R. Solli \etal showed that even in a passive transparent medium anomalous dispersion can be achieved by filtration of a component of a signal with normal dispersion \cite{solli03, mccormick03}. They  demonstrated this, theoretically and experimentally, for a strongly birefringent photonic crystal with electromagnetic waves of centimeter wave length. With an incoming wave that decomposes equally in a fast and a slow component and with an outgoing wave that is filtered in the same polarization direction as the incoming signal a dispersion relation was found that under half waveplate conditions could be interpreted as superluminal transit speed through the crystal. 
This result is interesting, since on one hand the anomalous dispersion of the filtered signal clearly is consistent with the interpretation of a superluminal group velocity and on the other hand it is clear that in this case a faster-than-light effect can only be apparent. Thus, the signal is created as a linear superposition of two waves, each of which are transmitted with subluminal velocity through the crystal. (In the context of quantum "weak value" measurements, the effect of the polarization filter can also be seen as performing a "postselection" of the photon state, as previously discussed in Ref. \cite{Ritchie91}.)

The results of ref.~\cite{solli03} is the motivation for the present work, where we examine how the derived group velocity is related to the transit of wave packets through the crystal. We first reproduce the earlier results by use of a  stationary phase analysis and focus then on how a signal of gaussian shape is moving through the crystal. The peak position of the outgoing signal is found to correspond to superluminal transit only when the wave packet is much wider than the crystal, and even in this case the filtered signal lies well within the envelopes of the two subluminal components, corresponding to the fast and slow wave components in the crystal.

\section*{Plane wave propagation and peak velocity}
\begin{figure}[h]
	\begin{center}
	\includegraphics[height=4cm]{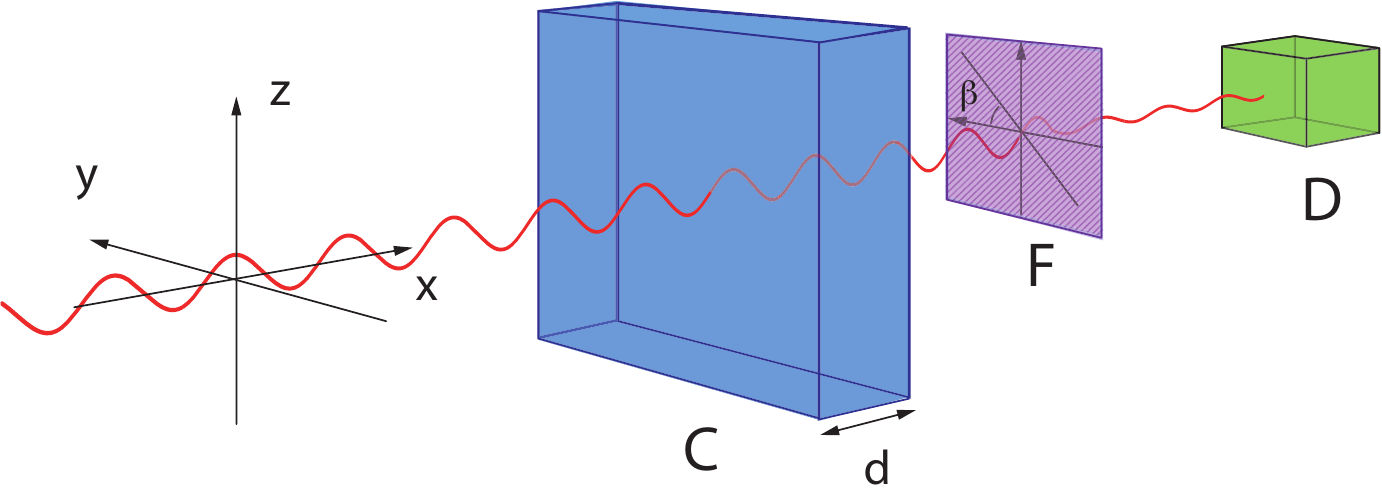}
	\end{center}
	\caption{Schematic illustration of the experimental situation considered in the text. A plane polarized electromagnetic wave is sent through a birefringent crystal $C$ of width $d$. After leaving the crystal the wave passes a polarization filter $F$ with a polarization angle $\beta$ that can be varied, and the amplitude of the filtered signal is registered in the detector $D$.}\label{krystall}
\end{figure}

The physical system we consider is sketched in Fig.~\ref{krystall}. A polarized electromagnetic wave is sent through a birefringent crystal, which is oriented with the optical axis ($z$-axis) 
orthogonal to the direction of propagation ($x$-axis). The boundary planes of the crystal, where the wave enters and leaves the crystal, are both assumed to be orthogonal to the direction of propagation. 
The outgoing wave passes a polarization filter so that the registered signal corresponds to linear polarization at a rotation angle $\beta$ in the y,z-plane. We consider here how the amplitude of the filtered signal depends on the angle $\beta$ and the wave number $k$ of the incoming wave.

The crystal is assumed to be linear and non-absorbing, with a dielectric tensor that is frequency independent.  Due to the difference between index of refraction $n_1$ outside  and $n_2$ inside the crystal, the reflection coefficient \cite{landau}
\begin{equation}
R=\left|\frac{n_1-n_2}{n_1+n_2}\right|^2,
\end{equation}
is slightly different from $0$. In fact, since the index of refraction is different for the ordinary wave ($n_2=n_o$), with polarization along the $y$-axis, and the extraordinary wave ($n_2=n_e$), with polarization along the $z$ axis, the reflection coefficient for these two directions of polarization are not identical. However, for both directions we will consider the deviation from $0$ to be sufficiently small so that the crystal can be treated as non-reflecting. (In Ref.\cite{mccormick03} the average index of refraction is given as about $1.25$ corresponding to a reflection coefficient $R\approx 0.01$.)
We shall in the following assume that $\Delta n\equiv n_e-n_o$ to be positive so that the ordinary wave is the fast and the extraordinary wave is the slow component in the birefringent material.

The incoming wave is assumed to be linearly polarized, with direction of polarization rotated $45 \textdegree$ relative to the $y$-axis. This means that the wave, inside the crystal, has components of ordinary and extraordinary waves of equal amplitudes. Due to the different propagation velocity of the two components  in the crystal, the polarization will change to elliptical polarization during the propagation. For a distance corresponding to {\em half-waveplate} condition the polarization is again linear, with polarization direction orthogonal to the polarization of the incoming wave. In the following we assume the length $d$ of the crystal in the direction of propagation to be close to that determined by the half waveplate condition $\Delta k\,d=(2N+1)\pi, N\in \mathbb Z$, with $\Delta k=k_e-k_o$ as the difference between the wave number of the extraordinary and ordinary wave. This means that the wave which leaves the crystal will be close to linearly polarized in the direction orthogonal to that of the incoming wave. The filtered signal for a polarization direction $\beta$ close to that of the incoming signal therefore picks out only a  small component of the outgoing wave, but this is the interesting one for the question of fast propagation through the crystal.

To analyze the situation, we start with the expression for the electric component of a monochromatic plane wave propagating through the crystal
\begin{equation}
\label{mono}
{\bf E(x,t)}=\frac{E_0}{\sqrt{2}}\begin{cases}
e^{i(kx-\omega t)}(\bs{\hat y}+\bs{\hat z}) & x<0 \\
e^{i(k_ox-\omega t)}\bs{\hat y}+e^{i(k_ex-\omega t)}\bs{\hat z}  & 0<x<d\\
e^{i(kx-\omega t)}(e^{i(k_o-k)d}{\bs{\hat y}+e^{i(k_e-k)d}\bs{\hat z}})  & d<x \,,
\end{cases}
\end{equation}
where $\bs{\hat  y}$ and $\bs{\hat  z}$ are the unit vectors in the y and z directions. The three intervals on the x-axis correspond to the regions where the wave is located before entering the crystal ($x<0$), when inside the crystal ($0 < x < d$) and after leaving the crystal ($d<x$). The polarization filter at this point has not been introduced. The wave numbers in the three regions are related to the frequency by
\be{}
 k=\frac{\omega}{c}, \quad k_o=\frac{\omega}{c}n_o, \quad k_e=\frac{\omega}{c}n_e \,,
 \ee
 with $n_o$ and $n_e$ as the indexes of refraction of the ordinary and extraordinary waves, respectively.
 
\begin{figure}[h]
	\begin{center}
	\includegraphics[height=6cm]{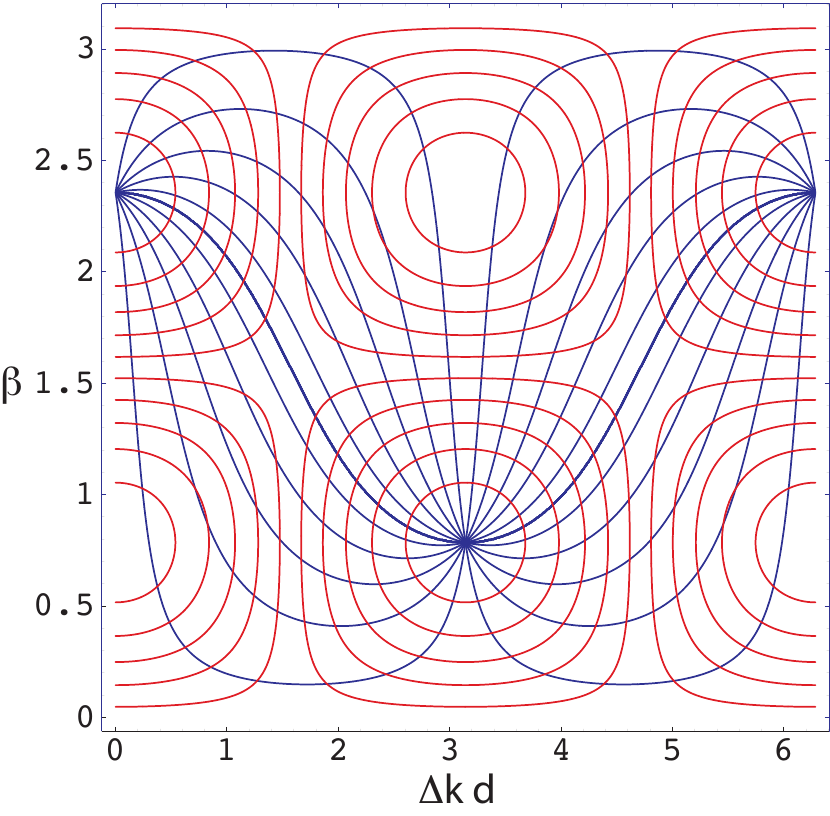}
	\end{center}
	\caption{Amplitude and phase of the filtered monochromatic plane wave as functions of the polarization angle $\beta$ and the wave number $k$. The (red) curves that circulate the singular points are contour lines of constant amplitude and the (blue) lines that interconnect these points are contour lines of constant phase. Three singular points with vanishing amplitude are shown, where the lower one corresponds to half waveplate condition for the transmitted wave. The numbers along the horizontal axis gives $\Delta k d$ measured from the closest point with an integer number of wave lengths ($2\pi N$). \label{konturplott}}
\end{figure}

 The filtered signal, with polarization at angle $\beta$ relative to the y-axis is then described by the electric field
 \begin{equation}
E_{\beta}={\bf E\cdot e}_{\beta}=\frac{E_0}{\sqrt{2}}e^{i(kx-\omega t)}e^{i(k_o-k)d}(\cos\beta+\sin\beta\, e^{i\Delta k d}) \,,
\end{equation}
with $\Delta k=k_e-k_o$.
We focus the attention on the  complex, relative amplitude
\be{}
 z=\halfsq(\cos\beta+\sin\beta \,e^{i\Delta k d})\equiv |z|e^{i\chi} \,,
 \ee
with absolute value and complex phase
 \begin{equation}
 \label{fase}
\begin{split}
|z|&=\sqrt{[1+\sin2\beta\,\cos(\Delta kd)]/2} \\
\chi &=\arctan\left(\frac{\sin(\Delta kd)}{\cos(\Delta kd)+\cot\beta}\right) \,.
\end{split}
\end{equation}
The amplitude $|z|$ vanishes at isolated points in the $(k,\beta)$-plane, determined by the conditions
\be{}
&a) \;\beta =\pi/4\; (5\pi/4),\quad \Delta k \,d = \pi\; ({\rm mod}\; 2\pi)\nn
&b)\; \beta =3\pi/4\; (7\pi/4),\quad \Delta k \,d = 0\; ({\rm mod}\; 2\pi) \,,
\ee
with a) corresponding to the half waveplate condition. The phase $\chi$ rotates by $2\pi$ around each point of zero amplitude, and the rapid variation close to this point is the reason for the large value of the group velocity near the half wave plate frequency. In Fig.~\ref{konturplott}  a graphic representation of the amplitude $z$ is shown as a function of $k$ and $\beta$ in the form of contour plots for $|z|$ and $\chi$.

The expression \pref{fase} for the phase $\chi$ of the filtered plane wave contains information not only about plane wave propagation in the crystal, but also, in an approximate sense, about propagation of wave packets, which in k-space are strongly localized around the wave number of the plane wave. Thus, a standard stationary phase analysis is based on the assumption that the peak position of the wave in x-space is determined by  the condition that the complex phase of the plane wave is stationary with respect to variations in $k$.

For the filtered wave the time dependent position of the peak is then given by
\begin{equation}\label{tid2}
\frac{d}{d\omega}(kx-\omega t +(k_o-k)d +\chi)=0 \Rightarrow x=ct+d-n_od-c\frac{d\chi}{d\omega} \,.
\end{equation}
Clearly the stationary phase argument identifies the group velocity, defined in the usual way $v_{group}=\dd{\omega}{k}$, as the velocity of the peak of the signal. However, when we consider the propagation of the wave after leaving the crystal there is no change relative to the velocity of light in vacuum. More interestingly, the expression gives the time delay (or time advance) relative to a wave that moves freely (without the presence of the crystal) \cite{Wigner55}. The shift in time of the outgoing, filtered signal relative to a freely propagating signal is
\begin{equation}
\label{tau}
\tau =\frac{d\chi}{d\omega}+\frac{(n_o-1)d}{c} \,,
\end{equation}
and with $\chi$ given by \pref{fase} we find
\be{tau2}
\tau&=&\frac{1+\cos(\Delta kd)\cot\beta}{1+\cot^2\beta+2\cos(\Delta kd)\cot\beta}\,\frac{d}{c}\,(n_e-n_o)+\frac{(n_o-1)d}{c}\nn
&=&\frac{d}{c}\left[\bar n -1-\half\, \frac{\cos2\beta\;\Delta n}{1+\cos(\Delta kd)\sin2\beta}\right]  \,,
\ee
where in the last expression we have introduced $\bar n=(n_e+n_o)/2$ and $\Delta n=n_e-n_o$.

\begin{figure}[h]
	\begin{center}
	\includegraphics[height=7cm]{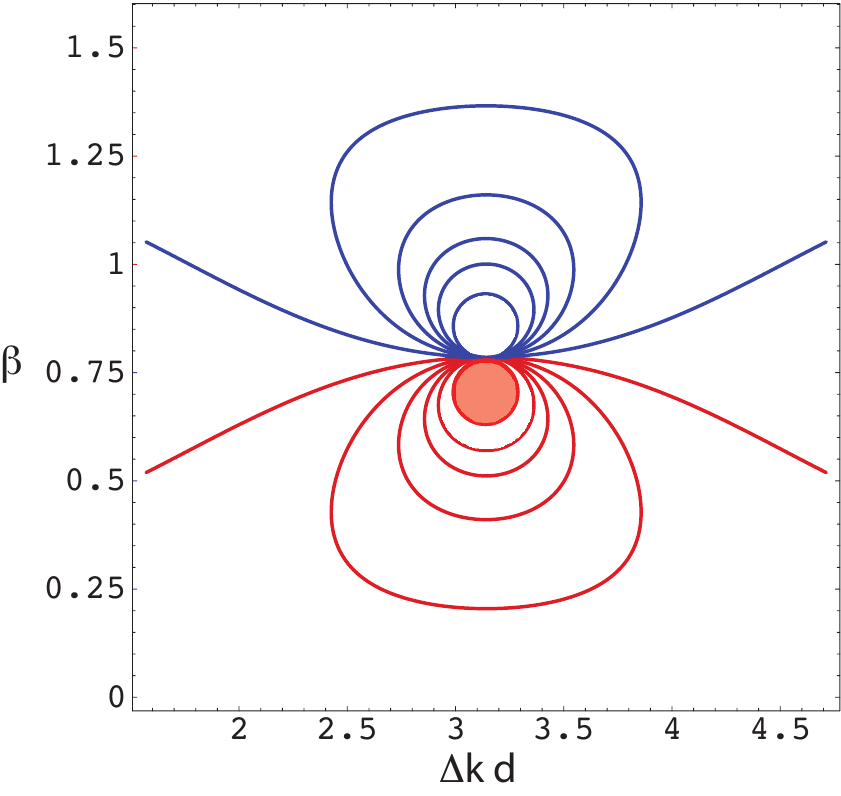}
	\end{center}
	\caption{Lines of constant time shift for the transmitted signal in the $\beta,k$-plane. The lower (red) curves correspond to advanced signals and the upper (blue) curves correspond to retarded signals. Close to the point of vanishing amplitude the time shift increases without any bound. The region of superluminal transit, for parameter values given in the text, is indicated by shaded red. \label{timeshift} }
\end{figure}
Eq.\pref{tau} shows that the time shift $\tau$ becomes negative when $\dd{\chi}
{\omega}$ is sufficiently large and negative,
\be{superlum}
\frac{d\chi}{d\omega}< -\frac{(n_o-1)d}{c} \,.
\ee 
This condition, which one may interpret as the condition for superluminal transit through the crystal, can be satisfied for parameter values close to the points of vanishing amplitude, $z=0$, where the phase angle is rapidly varying. To be more specific, we introduce new variables $\varphi$ and $\epsilon$,
\be{newangles}
\Delta k\, d = (2N+1)\pi+\varphi,\quad \beta=\frac{\pi}{4}+\epsilon \,,
\ee
and assume these to be sufficiently close to the point $\varphi=\epsilon=0$ so that an expansion to first order in these variables is sufficient,
\be{first order}
|z|=\half(4\epsilon^2+\varphi^2)^{\frac{1}{2}}, \qquad \chi=\mathrm{arctan}(\frac{\varphi}{2\epsilon})+\frac{\varphi}{2} \,.
\ee
This gives for the derivative of the phase $\chi$, 
\begin{equation}
\frac{d\chi}{d\omega}=(\frac{2\epsilon}{4\epsilon^2+\varphi^2}+\frac{1}{2})\frac{d\varphi}{d\omega}
=(\frac{2\epsilon}{4\epsilon^2+\varphi^2}+\frac{1}{2})\frac{d}{c}\Delta n \,,
\end{equation}
and the corresponding expression for the time shift of the peak, valid for small $\varphi$ and $\epsilon$,
\be{timeshift2}
\tau = {d\over c}\left(\bar n-1 +\frac{2\epsilon}{4\epsilon^2+\varphi^2}\,\Delta n\right) \,.
\ee
The expression shows that for negative $\epsilon$ and for sufficiently small $\epsilon^2$ and $\varphi^2$, the time shift $\tau$ may become negative. In fact, close to the singular point of the phase $\chi$ the time shift may become arbitrarily large and negative. Thus, the transit time may not only correspond to a superluminal propagation, but to a situation where the peak of the outgoing, filtered signal appears before the peak of the incoming signal has entered the crystal. Even if this situation may appear paradoxical, there is in reality no conflict with causality. The explanation is simply that the outgoing signal is being produced by the advanced tail of a sufficiently broad incoming wave.

The condition for superluminality \pref{superlum} for small $\epsilon$ and $\varphi$ reduces to the condition 
\begin{equation}\label{fulikhet}
\frac{(\epsilon+a)^2}{a^2}+\frac{\varphi^2}{4a^2}<1,\quad a\equiv \frac{\Delta n}{4(\bar n-1)} \,.
\end{equation}
 In Fig.~\ref{timeshift} the contour lines of constant time shift $\tau$ are shown around the point $\epsilon=\varphi=0$. The time shift is positive for $\beta > \pi /4 \;(\epsilon>0)$ and negative for $\beta < \pi /4 \;(\epsilon<0)$, with largest absolute values of $\tau$ close to the singular point $\epsilon=\varphi=0$. In the figure the region of superluminality is indicated by the shaded area for parameter value $a=0.15$.
 
The results discussed so far are in full agreement with those discussed in Ref.\cite{solli03}.  However, these results are based on the use of the stationary phase argument, which can only be trusted for a signal that is narrow in k-space and therefore wide in x-space. In the next section we shall therefore supplement this discussion by examining the propagation through the crystal of gaussian shaped electromagnetic waves of different widths.

\section*{Propagation of Gaussian shaped wave packets}
We introduce a wave packet with gaussian envelope in the direction of propagation by the following expression
\be{gauss}
\v E(x,t)=\sqrt{\frac{\sigma}{4\pi}}\int d\omega\, e^{-\frac{\sigma}{4}(\omega-\omega_0)^2} \v E(\omega;x,t) \,,
\ee
where $\v E(\omega;x,t)$ now denotes the monochromatic plane wave of Eq.~\pref{mono}.
The integral over frequencies is easily evaluated for each of the regions of propagation, and the expression for the $\beta$ component of the electric field is
\be{betafield}
E_{\beta}(x,t)&=&\frac{E_0}{\sqrt{2}}\, e^{-\frac{1}{\sigma c^2}(x-ct)^2}e^{i\frac{\sigma\omega_0}{4c}(x-ct)}(\cos\beta+\sin\beta)\,,  \quad\quad\quad\quad x<0 \nn
E_{\beta}(x,t)&=&\frac{E_0}{\sqrt{2}}\left[\cos\beta\, e^{-\frac{n_o^2}{\sigma c^2}(x-\frac{c}{n_o}t)^2}e^{i\frac{\omega_0n_o}{c}(x-\frac{c}{n_o}t)}\right. \nn
&&\left.  +\sin\beta\,e^{-\frac{n_e^2}{\sigma c^2}(x-\frac{c}{n_e}t)^2}e^{i\frac{\omega_0n_e}{c}(x-\frac{c}{n_e}t)}\right]\,,\quad\quad\quad\quad\quad  0<x<d \nn
E_{\beta}(x,t)&=&\frac{E_0}{\sqrt{2}}\left[\cos\beta\,e^{-\frac{1}{\sigma c^2}(x-ct-(1-n_o)d)^2}e^{i\frac{\omega_0}{c}(x-ct-(1-n_o)d)}\right.\nn
&&\left.+\sin\beta\, e^{-\frac{1}{\sigma c^2}(x-ct-(1-n_e)d)^2}e^{i\frac{\omega_0}{c}(x-ct-(1-n_e)d)}\right]\,,\quad  d<x \,.
\ee
Even if this component is filtered out only after the signal has left the crystal, it is instructive to study its propagation in all the three regions, and we show a plot of the shape of this component for a particular choice of parameter values below.

We first make a comparison between the peak position of the filtered gaussian signal and the corresponding position determined by the stationary phase argument. It is convenient to change to dimensionless variables by measuring length in units of the crystal width $d$, and as coordinate in the direction of propagation we choose
\be{xi}
\xi=\frac{x-ct}{d} +(\bar n-1) \,,
\ee
so that outside the crystal the propagating waves are stationary in this coordinate. $\xi=0$ corresponds to the position midway between the gaussians of the ordinary and extraordinary waves after leaving the crystal, while the the incoming gaussian wave  has the peak position
\be{speedoflight}
\xi_0=\bar n -1\,.
\ee
The peak position of the filtered signal, determined by the stationary phase argument, is under half waveplate conditions,
\be{statphase}
\xi_1&=&-\frac{c}{d}\tau +(\bar n-1)\nn
&=&\half\Delta n\,\frac{\cos\beta+\sin\beta}{\cos\beta-\sin\beta} \,,
\ee
and the condition for superluminality, as discussed earlier, is then expressed as $\xi_1>\xi_0$.

The true peak position of the time evolved gaussian signal can now be determined by use of Eq.~\pref{betafield}. For this signal we assume the central frequency $\omega_0$ to satisfy the half waveplate condition
\be{halfwave}
\frac {\omega_0}{c}\Delta  n\, d=(2N+1)\pi \,.
\ee
We consider the squared relative amplitude $f=(|E_{\beta} |/E_0)^2$ of the outgoing wave, which written as a function of $\xi$ has the form
\be{f-func}
f(\xi)=\half\left(\cos\beta\,\exp(-\mu^2(\xi-\half\Delta n)^2-\sin\beta\,\exp(-\mu^2(\xi+\half\Delta n)^2\right)^2 \,.
\ee
The maxima (and minima) of $f(\xi)$ are determined by the equation
\be{max}
\cos\beta\,(\xi-\half\Delta n)\,e^{\mu^2\xi\Delta n}-\sin\beta\,(\xi+\half\Delta n)\,e^{-\mu^2\xi\Delta n}=0 \,,
\ee
with
\be{mu}
\mu=\frac{d}{\ell}\,,\quad \ell=\sqrt{\sigma c^2} \,,
\ee
where $\ell$ measures the width of the incoming gaussian wave.

\begin{figure}[h!]
	\begin{center}
	\includegraphics[height=5cm]{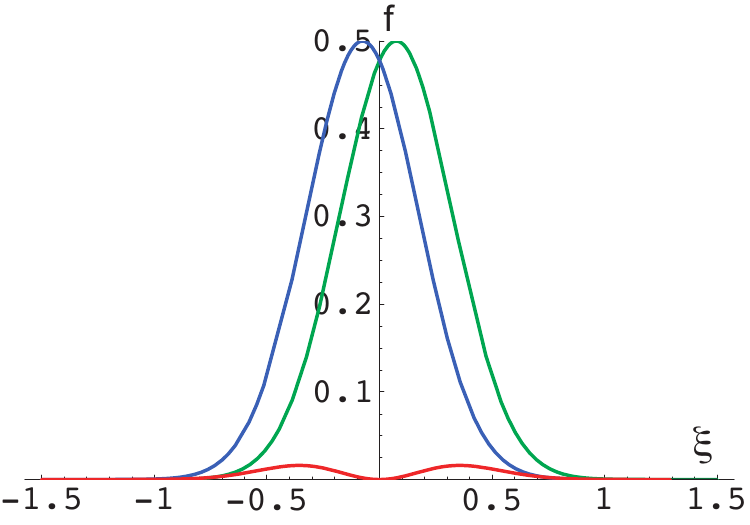}
	\end{center}
	\caption{Advanced and retarded signals. The figure shows the gaussian envelopes (the squared amplitudes $f$) of the fast (green) and slow (blue) components after leaving the crystal. For polarization angle $\beta=\pi/4$ there is destructive interference between the two components which gives rise to a symmetric signal with an advanced and a retarded maximum (red curve).\label{symmetric}}
\end{figure}

We note that Eq.~\pref{max} reproduces the solution $\xi=\xi_1$ in the limit $\mu^2\xi\Delta n\to 0$. This means that the inequality $\mu^2\xi\Delta n<<1$ has to be satisfied if the position determined by the stationary phase argument should be a good approximation to the real peak position of the outgoing signal. If we combine this condition with the condition for superluminality, $\xi_1>\xi_0$, we have the constraints
\be{constraint}
\ell>>\sqrt{\,\frac{\cos\beta+\sin\beta}{\cos\beta-\sin\beta}}\,\Delta n\, d>\sqrt{2\Delta n (\bar n-1)}\, d \,.
\ee
Thus,  the width $\ell$ of the signal has to be comparable to or larger than the width $d$ of the crystal in order for the time shift determined by the stationary phase argument to give a good approximation to the true time shift of the peak of the signal. Eq.~\pref{constraint} also shows that the closer $\beta$ is to the value $\pi/4$, the larger the width $\ell$ has to be. We further note from \pref{f-func} that the amplitude of the filtered signal decreases exponentially with  $\xi_1$ and therefore with the forward time shift $\tau$ of the signal. 

In reality $f(\xi)$ has two peaks rather than one, as follows from \pref{f-func}, since the function is defined by the partial cancellation of two shifted gaussians of opposite signs. This cancellation produces one advanced and one retarded peak. In Fig.~\ref{symmetric} this is shown for $\beta =\pi/4$. For  a wave with sharply defined value of $k$, this value of $\beta$ corresponds to the situation  where the $\beta$ component of the field vanishes. However, for a wave with a finite width in $k$-space there is only a partial cancellation between the fast (ordinary) and the slow (extraordinary) components. The resulting wave is symmetric with respect to the advanced and the retarded peaks, as shown in the figure. For values of $\beta$ smaller than $\pi/4$, there are still two peaks, but now there is an asymmetry, with the advanced peak  dominating the retarded one. 

\begin{figure}[h!]
	\begin{center}
	\includegraphics[height=5.5cm]{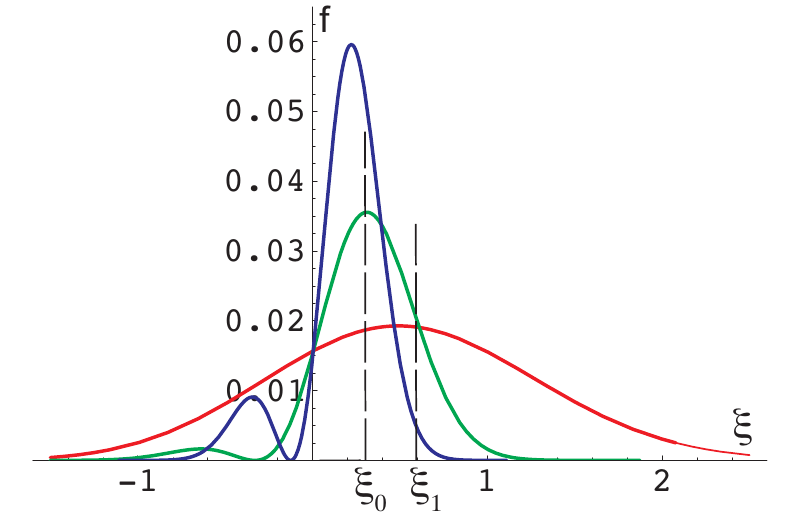}
	\end{center}
	\caption{Outgoing signals for gaussian input signals of different widths. The three curves show the squared amplitude $f$ as function of the dimensionless $x$-coordinate $\xi$ for polarization angle $\beta = 0.21 \pi$. For the (blue) curve with highest peak the width $\ell$ of the gaussian is given by $d/\ell=2.6$, with $d$ as the width of the crystal. The corresponding values are $d/\ell=1.6$ for the (green) curve of medium height and $d/\ell=0.6$ for the broadest (red) curve. The position $\xi_0$ indicates where the peak position would be if the gaussian signal moved with the speed of light, and $\xi_1$ if the signal moved with the speed determined by the stationary phase argument. These positions depend on the indexes of refraction which have here been chosen as  $\bar n=1.30$ and $\Delta n=0.15$.  \label{profiles}}
\end{figure}

In Fig.~\ref{profiles} this situation is illustrated by plots of the function $f(\xi)$ for the value $\beta=0.21 \pi$ and for three different values of the width $\ell$ of the gaussian, one larger  and two smaller than $d$. For the signal with smallest width the advanced and the retarded peaks are both clearly seen, however the amplitude of the retarded peak rapidly decreases with increasing width. One also notes that the forward shift of the advanced peak is close to the value $\xi_1$ for the broad signal with $\ell>d$, but is reduced below this value for the gaussians of smaller widths. 

\begin{figure}[h!]
	\begin{center}
	\includegraphics[height=11cm]{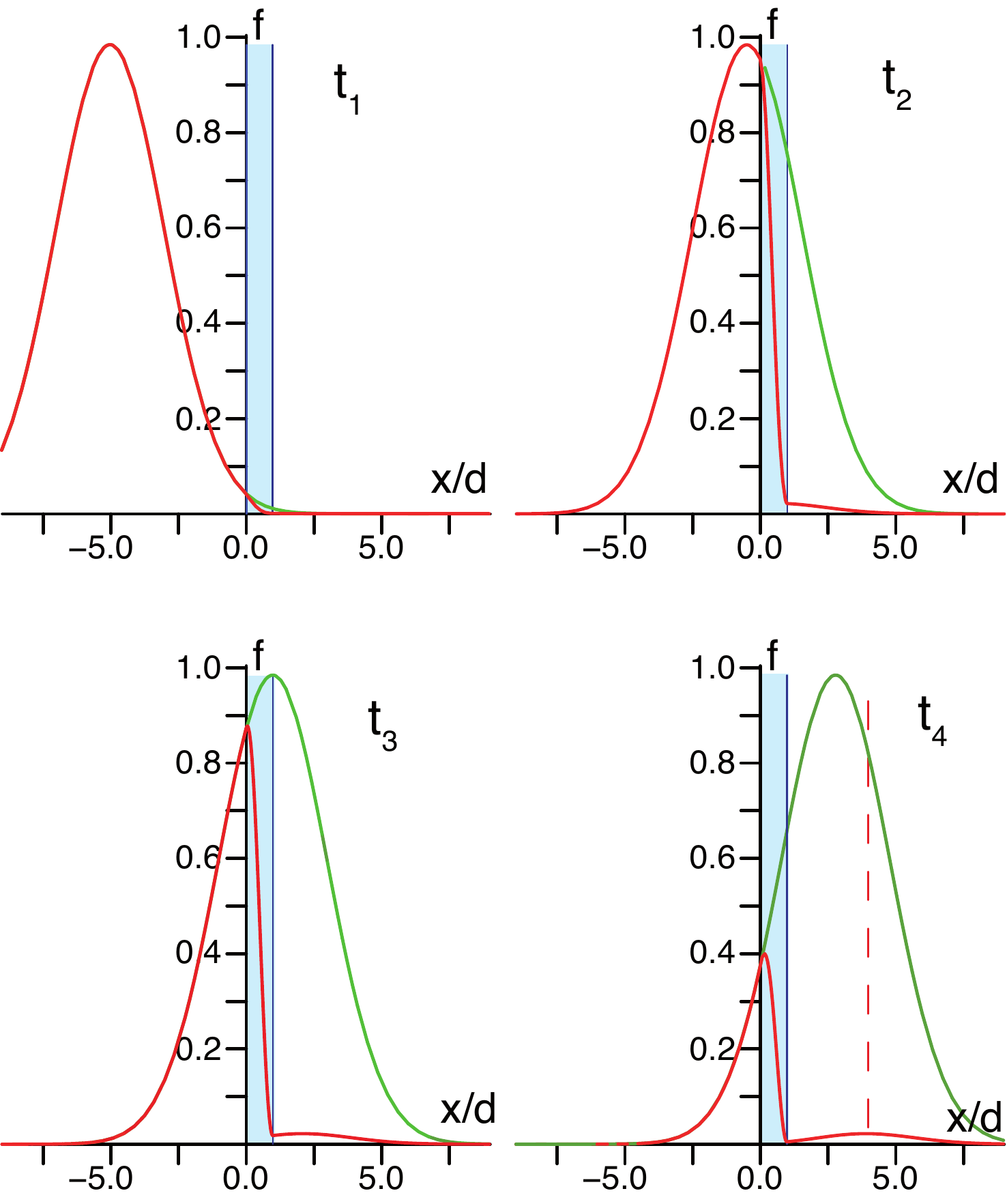}
	\end{center}
	\caption{Propagation of the $\beta$-component of the gaussian wave through the birefringent crystal. The figure shows the envelope of the component at four subsequent times, $t_1$, $t_2$, $t_3$ and $t_4$. The polarization angle is $\beta=0.21 \pi$ and the width is $\ell = 4 d$. The region where the crystal is located is indicated by the shaded (blue) area. The propagating wave of polarization $\beta$ is shown by the red curve, while the green curve is included for comparison to show the propagation of the same signal in empty space. The peak position of the transmitted signal, which is indicated by the dotted vertical line, clearly shows that the peak has an advanced position relative to the peak of the signal in empty space. In this plot the following values have been chosen for the indexes of refraction, $\bar n = 1.35$ and $\Delta n=0.5$. \label{evolution}}
\end{figure}
Finally, in Fig.~\ref{evolution} we show the time evolution of the $\beta$ component of the electric field, as the signal is transmitted through the crystal, in a case where the gaussian is sufficiently wide for the inequalities \pref{constraint} to be satisfied. The width of the gaussian in this case is, $\ell=4d$, and  value of the polarization angle is $\beta=0.21 \pi$. To enhance the time shift $\tau$ in this plot, the degree of birefringence is increased to $\Delta n=0.5$. The plot shows how the $\beta$ component is suppressed under the propagation in the crystal, so that only a small component survives and appears as the outgoing signal. The position of the peak of this signal can be seen to be advanced relative to a gaussian signal that propagates with the speed of light through the crystal.

\section*{Conclusion}
We have in this paper examined the phenomenon of superluminal propagation of electromagnetic waves through a birefringent crystal, an effect earlier discussed in Ref.~\cite{solli03}. The effect can be viewed as a two step process, where at the first step the crystal modifies the polarization of the propagating wave, due to different velocities of the ordinary and extraordinary components, but since there is no damping or reflection, the signal travels with subluminal velocity through the crystal.  However, at the next step, when the signal is sent through a polarization filter, the effect of the filter is to suppress the signal in such a way that the peak of the outgoing signal will appear as being shifted forward in time.

The amplitude of the filtered signal has been examined here as function of the frequency and polarization angle for an incoming monochromatic plane wave. Close to half waveplate condition, a stationary phase analysis indicates that the filtered signal may be substantially shifted forward in time, in agreement with a similar analysis in Ref.~\cite{solli03}. The implication for a modulated wave is that, for certain parameter values, the position of the peak of the filtered component will be shifted to an advanced position relative to a freely propagating wave. This is what in \cite{solli03} is interpreted as a superluminal transit of the signal through the crystal.

To examine this effect further we have considered incoming waves with gaussian envelopes of varying widths.  The effect of the filter on the transmitted wave will in general be to reduce the signal in such a way that a two-peaked signal will appear, with an advanced and a retarded part. For parameter values corresponding to superluminal transit time the advanced peak will be the dominating one. However, a good agreement between the stationary phase analysis of the peak position and the true peak position of the gaussian signal depends on a sufficiently broad gaussian envelope in the direction of propagation. We find that if the width of the signal is small compared to the width of the crystal, the peak will be located behind the position determined from stationary phase analysis and the filtered wave in this case will also show both the advanced and retarded peaks. If instead the width of the gaussian is much larger than the crystal width, the peak position will be close to that determined from stationary phase and it may be substantially advanced relative to signal moving with the speed of light. However, the broadness of the signal means that it is not strongly located around this position.

The stationary phase analysis shows that the situation discussed here is similar to other cases, where reduced transit time can be seen as a consequence of the anomalous dispersion of the transmitted signal. A particular such case is quantum tunneling, where the outgoing wave may appear with reduced amplitude but advanced position \cite{hartman62,hauge89,Chiao97}. However, the present case is, in a sense, more transparent than the other cases due to the separation of the effect into two parts. Thus, at the first step, when the wave propagates through the crystal, there is no faster-than-light propagation, but only a change of polarization. At the next step, the polarization filter produces the advanced signal by separating out a special superposition of the two subluminal waves. That means that  the superluminal effect, in the present case, can be viewed as due to fast transit of interference patterns through the crystal. Viewed in this way the effect seems close to other cases where velocities larger than $c$ can be obtained by interference.  A  particular case is the superluminal X-wave in free Maxwell theory, which is created by interference between monochromatic plane waves \cite{Rodrigues97,Saari97}. \\

\noindent
{\bf Acknowledgment}\\
We would like to thank Prof. Raymond Chiao for his useful comments. This work has financial support from the Research Council of Norway and from NordForsk.



\begin{thebibliography}{99}

\bibitem{sommerfeld14}
A. Sommerfeld,
{\em {\"U}ber die fortpflanzungdes lichtes in dispergierenden medien},
Ann. Phys. {\bf 44}, 177(1914).

\bibitem{brillouin14}
L. Brillouin,
{\em {\"U}ber die fortpflanzungdes lichtes in dispergierenden medien},
Ann. Phys. {\bf 44}, 203(1914).

\bibitem{brillouin}
L. Brillouin,
{\em Wave Propagation and Group Velocity},
(Academic Press, New York, 1960).

\bibitem{Chu82}
S. Chu and S. Wong,
{\em Linear pulse propagation in an absorbing medium}, 
Phys. Rev. Lett. {\bf 48}, 738 (1982)

 \bibitem{Segard85}
 B. Segard and B. Macke,
{\em Observation of negative velocity pulse propagation}, 
Phys. Lett. A {\bf 109}, 213 (1985).

\bibitem{Steinberg93}
A. M. Steinberg, P. G. Kwiat, and R. Y. Chiao,
{\em Measurement of the Single-Photon Tunneling Time},  
Phys. Rev. Lett. {\bf 71}, 708 (1993).

\bibitem{Wang00}
L. J. Wang, A. Kuzmich, and A. Dogariu,
{\em Gain-assisted superluminal light propagation}, 
Nature {\bf 406}, 277 (2000).

\bibitem{Chiao93}
R.Y. Chiao,
{\em Superluminal (but causal) propagation of wave packets in transparent media with inverted atomic population}, Phys. Rev. A {\bf 48} R34 (1993).

\bibitem{Chiao96}
R.Y. Chiao,
{\em Population inversion and superluminality} in {\em Amazing light: A volume dedicated to Charles Hard Townes on his 80th birthday}, edited by R.Y. Chiao (Springer Verlag, New York, 1996), p.91.

\bibitem{solli03}
D. R. Solli, C. F. McCormick, C. Ropers, J. J. Morehead, R. Y. Chiao, J. M. Hickman,
{\em Demonstration of superluminal effects in an absorptionless, non-reflective system},
Phys. Rev. Lett., {\bf 91}, 143906 (2003),\\
D. R. Solli,  C. F. McCormick,  R. Y. Chiao and  J. M. Hickmann,
{\em Experimental observation of superluminal group velocities in bulk two-dimensional photonic bandgap crystals},
 Journal of Selected Topics in Quantum Electronics, {\bf 9}, 40 (2003).
 
 \bibitem{mccormick03}
D. R. Solli, C. F. McCormick, R. Y. Chiao, J. M. Hickman,
{\em Birefringence in two-dimensional bulk photonic crystals applied to the construction of quarter waveplates},
Optics express {\bf 11} 125 (2003).

\bibitem{Ritchie91}
N.W.M. Ritchie, J.G. Story and Randall G. Hulet,
{\em Realization of a Measurement of a "Weak Value"},
Phys. Rev. Lett., {\bf 66}, 1107 (1991).

\bibitem{landau}
L. D. Landau, E. M. Lifshitz and L. P. Pitaevskii,
{\em Electrodynamics of Continuous Media},
 Pergamon Press, Second Edition 1984.
 
 \bibitem{Wigner55}
 E.P.Wigner,
 {\em Lower limit for the energy derivative of the scattering phase shift},
 Phys. Rev. {\bf 98}, 145 (1955).
 

\bibitem{hartman62}
T.E. Hartman,
{\em Tunneling of a wave packet},
J. Appl. Phys. {\bf 33}, 3427 (1962).

\bibitem{hauge89}
E.H. Hauge and J.A. St{\o}vneng,
{\em Tunneling times: a critical review},
Rev. Mod. Phys, {\bf 61}, 917 (1989).

\bibitem{Chiao97}
R. Y. Chiao and A. M. Steinberg, {\em Tunneling Times and Superluminality} , in 
Progress in Optics, edited by E. Wolf, Vol. 37, (Elsevier, Amsterdam, 1997), p. 347.

\bibitem{Rodrigues97}
W.A. Rodrigues and J.Y. Lu, "
{\em On the Existence of Undistorted Progressive Waves (UPWs) of Arbitrary Speeds in Nature},
 Found. Phys. {\bf 27}, 435 (1997).

\bibitem{Saari97}
P. Saari, and K. Reivelt,  
{\em Evidence of X-Shaped Propagation-Invariant Localized Light Waves},
 Phys. Rev. Lett. {\bf 79}, 4135 )1997).

\end{thebibliography}
\end{document}